# COVID-19 risk-perception in long-distance travel

*Nejc Geržinič, Maurizio van Dalen, Oded Cats*

## Abstract

Long-distance / International travel is a topic within travel behaviour research that has seen little attention in the past, largely due to the irregular and sporadic nature of such trips. And yet, a single long-distance trip can amount to a distance equivalent to a year's worth of commute trips, resulting in a similar, if not worse, environmental footprint. Understanding travellers' behaviour is therefore just as relevant for such trips, as it is for everyday commute trips. As international travel is slowly picking up from the COVID-19 pandemic, it has been marred by an abundance of national and regional pandemic-related safety measures. While their primary goal is to protect the local population from infection, these safety may also make travellers feel safer while travelling. This perceived safety can – and likely does – differ from the true efficacy of the measures. In this research, we investigate people's perception of eight COVID-19-related safety measures related to long-distance trips and how subjective perception of safety impacts their mode choice among car, train and aircraft. We employ a Hierarchical Information Integration (HII) approach to capture subjective perceptions and then model the obtained data by means of a Latent Class Choice Model, resulting in four distinct segments. To extrapolate the segments onto the rating experiment of HII, we apply a weighted least squares (WLS) regression, to obtain segment-specific safety perception. Two segments show a relatively high value-of-time (72€/h and 50€/h), tend to be more mode-agnostic and prefer determining the level of risk by themselves (relying primarily on infection and vaccination rate). The remaining two segments have a lower value-of-time (38€/h and 15€/h) and have strong mode affinity, for the train and car respectively. Future research could look into a way that segments the sample based on both the mode choice and rating experiment, providing additional insights into the heterogeneity of individuals in their perceptions.

## 1   Introduction

In recent years, long-distance travel has been gaining prominence in both political and scientific discussions (particularly in Europe), with many new services, proposals and policies being passed or put forward, aimed at fundamentally reshaping how we travel (euronews, 2023; Steer, 2021; Witlox et al., 2022). What ties them all together is their emphasis on improving rail travel. In recent years, night trains have been experiencing a revival in Europe (de Kemmeter, 2021; Heufke Kantelaar et al., 2022). France and Austria have banned short-haul domestic flights on routes with good (high-speed) rail alternatives (Ledsom, 2022; Morgan, 2020). Proposals from several organisations have also been put forward for Europe-wide (high-speed) rail infrastructure and service enhancement, with the most high-profile case being the TEE 2.0 project (German Federal Ministry of Transport and Digital Infrastructure, 2021).

To support the evaluation of proposed policies and investment projects, there has been a fair amount of research on long-distance travel and high-speed rail. An extensive literature review on research concerning long-distance travel is given by Sun et al. (2017), providing a good overview of the state-of-the-art. As most studies have a specific case-study focus, it is difficult to obtain a clear and straightforward understanding of individuals' long-distance travel behaviour. The consequences of introducing high-speed rail (HSR) differ greatly per country, based on the implementation, service pattern, policies and regulation of the air, rail and road markets etc. Jiang et al. (2021) acknowledge that trains are certainly a more sustainable alternative, but due to induced demand, the environmental benefits may not be as high as anticipated.

With respect to long-distance travel behaviour, most studies (both revealed (RP) and stated preference (SP)) report a Willingness-to-Pay (WtP) for in-vehicle time and access/egress time to HSR train stations, with the former being in the range 10-30 €/h and the latter at 20-50 €/h (Bergantino & Madio, 2018; Ortúzar & Simonetti, 2008; Román et al., 2014; Román & Martín, 2010). Other frequently evaluated attributes are frequency/headway, waiting time, reliability and comfort, with the results between studies being highly inconsistent. Frequency and waiting time essentially analyse the same travel aspect, but the perception of this is fundamentally different in long-distance travel and thus cannot be compared to values of waiting time reported in studies on daily commute behaviour. The context-specific nature of the studies (Spain, Italy, Chile) further exacerbates these differences.

*Hierarchical information integration*

In addition to the most frequently evaluated attributes, people consider many other aspects when making decision about long-distance travel which can be difficult to capture, or there may simply be too many to consider. To overcome this issue, some studies employ the Hierarchical Information Integration (HII) approach. It enables the analyst to capture a broader array of attributes and grouping them based on a common denominator. Respondents first evaluate these groups of attributes (i.e. comfort, safety, reliability, convenience,…) individually, giving them a subjective ranking on a Likert-scale. These rankings are then presented to respondents in a bridging experiment, which is able to couple the different attributes and disentangle how they are traded-off (Louviere, 1984). Since its introduction, different version of HII have been proposed and implemented. Molin & Timmermans (2009) present a review of HII studies and classify the studies into three categories: (1) Conventional HII, as defined by Louviere (1984), (2) HII variant that was first introduced by Bos et al. (2004) and (3) an Integrated Choice experiment, proposed by Oppewal et al. (1994). The HII variant differs from the main approach in the setup of the bridging experiment, with some attributes representing the subjective construct evaluations, alongside conventional objective attributes such as travel time or cost. The Integrated Choice experiment on the other hand, includes attributes of other constructs in the sub-experiments as well, giving more context to the decisions being made. Two notable examples that utilise HII to analyse long-distance travel are the study by Molin et al. (2017), focusing on the perception of airline safety, following the crash of Malaysia Airline Flight MH17 in eastern Ukraine; and the work of Heufke Kantelaar et al. (2022) that looked into the perception of comfort on night trains, stemming from the re-popularisation of night rail travel in Europe. Both studies used the *HII variant* approach, including a single construct for safety and comfort, respectively, with the other attributes being included in the bridging experiment.

Both above mentioned studies (Heufke Kantelaar et al., 2022; Molin et al., 2017) modelled the bridging experiment by means of a Panel Mixed Logit (ML) model, which has the benefit of capturing respondent heterogeneity, as well as being able to account for the panel effect of the SP data. An alternative approach to capturing respondent heterogeneity is the latent class choice model (LCCM) (Greene & Hensher, 2003). Where the ML model assumes a distribution of parameters in the population, the LCCM estimates separate multinomial logit (MNL) models for each class, meaning they are easily distinguishable and their interpretation is more clear-cut. Another benefit of LCCMs is the incorporation of socio-demographic information in the class allocation function, giving information of the class compositions. However, applying LCCMs to HII data is challenging, as the perception of the subjective ratings in the bridging experiment will result in different parameters for each class. However, if the rating experiment is modelled with a regression function as is common practice now, that means there is no distinction between classes in their perception of the different influencing attributes. Intuitively, this subjective perception should differ between the classes, yet to the best of the authors' knowledge, latent class segmentation has not been attempted in HII variant experiments.

*Covid-19 pandemic*

With the outbreak of the covid-19 epidemic, the perception of safety and infection risk has become a key decision factor for many travellers. Since the outbreak in late 2019, our travel behaviour has fundamentally altered. The most notable change happened during the first lockdown, when all but essential activities were cancelled, leading to drastic reductions in travel demand, with public transport (PT) seeing the biggest drop in usage (Currie et al., 2021; de Haas et al., 2020; Shamshiripour et al., 2020).

According to Currie et al. (2021), the fear of infection remained a key factor for travellers to continue avoiding the PT in subsequent lockdowns. Crowding was also high on the list of influential factors, due to the higher likelihood of virus spread in crowds. To combat the spread of covid-19, numerous measures were introduced on PT around the world (Shelat et al., 2022; Shortall et al., 2022; Tirachini & Cats, 2020), such as enhanced cleaning policies, increased ventilation, mask mandates, travel and country entry regulations, adapted operating strategies etc.

Long-distance and particularly international travel were especially strongly impacted by the outbreak of the epidemic, with most countries implementing strict entry requirements or closing their borders entirely. This meant that most international trips had to be cancelled or rescheduled if possible (Fatmi et al., 2021; Mary & Pour, 2022). Due to both the safety perception (infection risk) and a reduced level of service, mode choice was affected (Li et al., 2021), with more people choosing to travel by car (Kamplimath et al., 2021; Shamshiripour et al., 2020) rather than train or air. Similar to what was reported for commute behaviour by Currie et al. (2021), hygiene became a top priority for individuals when selecting their travel mode for long-distance trips (Kamplimath et al., 2021). In terms of future prospects, researchers propose mixed outcomes, with some studies reporting people flying less after the pandemic (de Haas et al., 2020; Shamshiripour et al., 2020), with the shift being mainly towards the private car (Shamshiripour et al., 2020), whereas others reported the train being perceived as less safe than flying, with the latter being perceived as no less safe than the car (Kamplimath et al., 2021). Nevertheless, Burroughs (2020) and Tardivo et al. (2020) both speculate that, particularly in Europe, the railway sector could come out of the covid-19 epidemic far stronger. They attribute this in part due to the ever more important environmental concerns of society, large investments into railways during the epidemic in the form of economic relief packages and partially due to lower risk of infection on trains as opposed to aircraft.

The impact of the covid-19 epidemic on long-distance travel and the associated behaviour of individuals is therefore uncertain. Despite a large array of measures being passed, little is known about the subjective perception of their efficacy. Although policymakers highlight the importance of measures and their benefits, it is ultimately the perceived efficacy which underlies users' decision making. To the best of our knowledge, this has not yet been looked into with respect to the long-distance travel market. And although the covid-19 epidemic seems to slowly be fading in large parts of the world, research suggests that future pandemics are becoming increasingly likely (Marani et al., 2021; Michael Penn, 2021), meaning that our understanding of the perception of public safety is just as important now, if not more, than during an ongoing pandemic.

*Paper contribution and outline*

The contributions of this paper are twofold. Firstly, we evaluate the perception of various covid-19 measures aimed at limiting the spread of the virus, through an HII variant type SP survey. The rating experiment includes eight attributes associated with the perception of infection risk. This infection risk is then carried into the bridging experiment, along with travel cost, time and comfort level, where respondents choose their preferred travel mode for a long-distance trip of approximately 500km and 1000km. Secondly, upon modelling the bridging experiment by means of an LCCM, we estimate several weighted least squares (WLS) regression models to uncover the different perceptions of infection risk as experienced by different population segments that are obtained from the LCCM. Using a WLS regression, as opposed to the commonly applied ordinary least squares (OLS) regression, the analyst is able to obtain segment-specific perceptions from the rating experiment of HII for any analysed context, providing more information on how the perception of a construct differs among individuals.

The remainder of the paper is structured as follows. The survey design, including the risk-of-infection perception experiment and the bridging mode choice experiment are presented in Section 2.1. Model estimation information for the WLS regression and LCCM are defined in Section 2.2, with the data collection and relevant covid-19 situation outlined in Sections 2.3 and 2.4, respectively. The results of the final model are presented in Section 3, with the implications, application and trade-off behaviour showcased in Section 0. Finally, the research is summed up, with the limitations and outlook for future research discussed in Section 5.

## 2 Methodology

In this section, we outline the steps undertaken in this research: how the survey was constructed and thereafter, the obtained data modelled. As the HII experiment is formed of two separate experiments (the rating and bridging experiments), they are each discussed separately, both in the survey design section (2.1) as well as in the model estimation section (2.2). Section 2.3 then presents the data collection approach and the sample characteristics, with Section 2.4 outlining the covid-19 circumstances under which respondents were filling in the survey.

### 2.1 Survey design

As outlined in the Introduction, safety perception with respect to covid-19 infection risk is arguably a complex construct, associated with a large number of possible influencing factors. To capture their impact on individuals' mode choice, we devise an HII experiment, specifically the HII variant, developed by Bos et al. (2004) which has been applied in past studies by Molin et al. (2017) and Heufke Kantelaar et al. (2022). We design two separate experiments: (1) a rating experiment that captures respondents' perception of infection risk and (2) a bridging experiment, where the infection risk rating is included as one of the attributes. The experiments are administered sequentially, starting with the rating experiment, where the respondents get acquainted with the infection ratings and different attribute levels, subsequently followed by the bridging experiment where those subjective ratings are contextualised in a full mode-choice experiment. The design of each experiment is described in more detail in the following sections.

*Rating experiment*

In the rating experiment, attributes pertaining to a common topic are joined and their attributes are varied in order to obtain the influence of each individual factor onto the overall perception of that construct. A single rating experiment is administered in our experiment, where respondents report their perceived risk of infection with covid-19 in relation to a long-distance trip.

Individuals may consider a wide variety of factors and mitigation policies when evaluating their perceived risk of infection. Numerous studies measured reported perceived risk directly (asking respondents about their risk perception) to understand its relation to other factors (Dryhurst et al., 2020; Kroesen et al., 2022; Mertens et al., 2020). However, there is lack of knowledge concerning the underlying determinants of the perceived risk. An SP survey on mode-choice in Santiago, Chile was carried out by Basnak et al. (2022), in which the authors test respondents' sensitivity to mask-wearing compliance (% of passengers wearing a mask), crowding and cleaning policy. Utilising an HII experiment, Shelat et al. (2022) analysed infection risk perception and its impact on train route choice in the Netherlands. In the risk perception experiment, they tested on-board crowding, number of transfers, face mask policy, sanitisation, current infection rate and lockdown status. Crowding has also been recognised as a major influencing factor on mode choice by Currie et al. (2021). None of the above has been conducted in the context of long-distance travel, where train and aircraft are the main passenger transport alternatives.

We devise an experiment with eight attributes, based on three groups as defined by Shelat et al. (2022): *trip-specific* (partial control by the operator), *policy-based* (set by the government or operator) and *pandemic-context* (information on the state of affairs at the time). The attributes and associated attribute levels are summarized in Table 1.

Regarding the *trip-specific* attributes, we include **on-board crowding**, which is one the most frequently cited influencing factor on the risk perception (Basnak et al., 2022; Currie et al., 2021; Shelat et al., 2022). As standing is a rare occurrence on long-distance travel and in many cases not even permitted, we include this as a share of occupied seats in the vehicle.

With regard to *policy-based* attributes, we include **face mask policy**, which is one of the more recognisable policies adopted by governments and operators around the world (Shortall et al., 2022; Tirachini & Cats, 2020). We also test for **cleaning policy** and **air circulation**. Although the efficacy of the former is contested (Thompson, 2020), it may still have a profound impact on travellers risk perception (Basnak et al., 2022), and it is therefore included. Many airlines were quick to emphasize their commitment to hygiene and have put out statements on their enhanced cleaning policies and the use of HEPA filters in air-conditioning units (Wichter, 2020). Ventilation has also been a frequent piece of advice to the public as an easy and efficient way of reducing infection risk.

We test two more policy-based attributes, which are specifically aimed at international travel: **government travel advice** and **entry requirements**. In the Dutch context, long-distance travel mostly implies international travel. The Dutch government keeps a regularly updated list of countries and a simple colour-coded travel advice (green, yellow, orange, red) for each country, based on the risk associated with traveling there (Rijksoverheid, n.d.). During the pandemic, this list was updated given the epidemiological situation at the time (case numbers, local regulations etc.). For many travellers, it is a first point of information and a good indication of the associated risks. At the time of the data collection (February and March, 2022), testing and vaccination had already become widespread across Europe and with the introduction of the QR-code system (European Commission, 2021), many countries adopted this as a means to allow for some international travel while keeping with national containment policies. Depending on the government policies and the severity of the pandemic at the time, different combinations of certificates could be required to enter a country (vaccination, recovery, testing).

Finally, the *pandemic situation* at the decision moment is an important consideration individuals make. Initially, this primarily meant the **infection rate** (Shelat et al., 2022), with the most frequent metric being number of cases (although hospitalisations, ICU admittance or virus reproduction rate have also been reported by governments). With vaccination becoming more widespread and the concept of herd immunity, the **vaccination rate** in a society can also be a predictor of overall infection risk perception.

As prior values are not available for some of the included attributes, an orthogonal (fractional factorial) design is utilised to construct the experiment. The resulting design contains 20 rows. By applying blocking, each individual was asked to evaluate five choice sets, indicating their perceived level of infection risk on a Likert-scale between 1 and 5. The design is obtained by utilising Ngene software (ChoiceMetrics, 2018) and the full design can be seen in Table 7 in Appendix A.

*Table 1. Attributes and attribute levels used in infection risk rating experiment*

| Category | Risk factor | Attribute levels |
|---|---|---|
| Trip-specific | **On-board crowding** | - 25% of seats occupied<br>- 50% of seats occupied<br>- 75% of seats occupied<br>- 100% of seats occupied |
| Policy-based for travel | **Face mask policy** | - No mask mandatory<br>- Mandatory face mask (can be any kind)<br>- Mandatory surgical face mask<br>- Mandatory FFP2 mask |
| Policy-based for travel | **Air circulation** | - No ventilation or air-conditioning<br>- Only ventilation<br>- Air-conditioning without HEPA filters<br>- Air-conditioning with HEPA filters |
| Policy-based for travel | **Cleaning policy** | - The same cleaning policy as before covid-19<br>- Enhanced cleaning (touch points)<br>- Weekly full-vehicle disinfection<br>- Daily full-vehicle disinfection |

| Policy-based for international travel | **Travel advice** | - Green<br>- Yellow<br>- Orange<br>- Red |
|---|---|---|
| Policy-based for international travel | **Entry requirements** | - No entry regulations<br>- Tested, recovered or vaccinated (3G)<br>- Vaccinated or recovered (2G)<br>- Vaccinated or recovered + tested (2G+) |
| Pandemic-context | **Infection rate** | - 100 positive tests per day (summer 2020 & June 2021)<br>- 10.000 positive tests per day (autumn 2020 & July 2021)<br>- 25.000 positive tests per day (November 2021)<br>- 100.000 positive tests per day (fictitious extreme high) |
| Pandemic-context | **Vaccination rate** | - 15% fully vaccinated (Bulgaria)<br>- 30% fully vaccinated (Romania)<br>- 70% fully vaccinated (Netherlands & EU average)<br>- 90% fully vaccinated (Portugal) |

*Bridging experiment*

To link the perceived infection risk with other travel-related attributes, a bridging experiment is designed. Based on the HII variant (Bos et al., 2004), this experiment contains both the rating experiment attribute value, and directly included objective attributes. This bridging experiment is designed as a mode choice experiment, wherein respondents can choose among car, train or aircraft options, as these are the most widely available and have also seen most attention in research (Bergantino & Madio, 2018; Cascetta et al., 2011; Ortúzar & Simonetti, 2008; Pagliara et al., 2012; Román et al., 2014; Román & Martín, 2010). Despite the growth of long-distance bus services (such as Flixbus) in Europe in the years before the pandemic, those are still primarily seen as a low-cost alternative, and are therefore excluded from the survey. The aircraft alternative is defined as a "flag carrier", to avoid respondents making assumptions on the type of service offered. For train, no distinction is made between conventional rail and high-speed rail, as the interior comfort level is often indistinguishable and travel time is the only indicator of the travel speed.

The most important attributes, included in past research are **travel time** and **cost**. We define travel time as the door-to-door travel time, including the main leg travel time, the terminal dwell time (time spent at the airport/train station) and the access/egress time. The latter is often included in studies, because long-distance/international trains serve only a single or a handful of stations in a city, resulting in a significant role of access/egress time in the decision-making process. This is even more pronounced for airports, as they tend to be located outside of the city, sometimes far away, resulting in long access times. Different travel time components are merged into a single attribute to avoid overwhelming respondents with too many attributes. Transfers are not included, as in long-distance travel, they tend to be highly case dependent. Frequency and time-of-day information is also left out, to minimise the amount of information that respondents need to process and evaluate.

In addition to the **perceived infection risk**, the fourth and final attribute included in the survey is **travel class (comfort level)**. The travel class can have a strong impact on the perception of travel time and with more personal space and often lower occupancy in first/business class, some may choose it as a safer travel alternative.

To capture a broader scope of potential long-distance trips, two separate experimental designs, with two distance categories are used: a shorter trip of approximately 500km and a longer trip of approximately 1000km. These distances are used primarily to determine appropriate travel time and travel cost attribute levels. The attribute levels for all three modes in both distance categories can be viewed in Table 2. Example destinations from Amsterdam are also given to respondents as an indication of the travel distance:

- 500km trip:   London, Paris, Zurich, Berlin, Copenhagen
- 1000km trip:  Bordeaux, Barcelona, Milan, Warsaw, Stockholm

A Bayesian D-efficient design in Ngene (ChoiceMetrics, 2018) is generated for the bridging experiment. An advantage of an efficient design is that it results in far fewer choice sets. Using the approximate willingness-to-pay (WtP), the design maximises the number of choice tasks within this trade-off area and avoids dominant alternatives. As we can never be fully certain about the priors, especially when stemming from different sources, we apply a Bayesian efficient design. This allows us to specify a standard error for each prior value, indicating our level of certainty. The priors for the travel time and cost are based on the study by Kouwenhoven et al. (2014), who carried out a detailed value-of-time (VOT) study for the Dutch Government. The values tend to be around 10 €/h, which we use as a base. For risk perception, we set the prior at a WtP of €5 per risk level reduction, based on the result of Shelat et al. (2022), who found a value of ~€4 per risk level reduction for trips up to one hour long. Finally, we use a WtP for a higher level comfort (business/first class) of €50, based on the findings of Ortúzar & Simonetti (2008) and also on the values observed when determining the price levels. The standard errors of the priors are set at half the value of the prior. Given the assumed normal distribution, this means that we are 0.975 certain that the prior has the correct sign (negative for travel time, cost and perceived infection risk & positive for comfort). The priors and their respective standard errors can be found in Table 2. The final designs for both distance scenarios are presented in Table 8 and Table 9 in Appendix A.

*Table 2. Prior parameter values, attributes and attribute levels per mode and distance category*

| | Prior values | Train | | Aircraft | | Car | |
|---|---|---|---|---|---|---|---|
| | | *~500km* | *~1000km* | *~500km* | *~1000km* | *~500km* | *~1000km* |
| **Travel time** | -0.1 (0.05) | - 3h<br>- 4.5h<br>- 6h | - 6h<br>- 9h<br>- 12h | - 3h<br>- 4h<br>- 5h | - 4h<br>- 5h<br>- 6h | - 4.5h<br>- 6.5h<br>- 8.5h | - 10h<br>- 13h<br>- 16h |
| **Travel cost** | -0.01 (0.005) | - €30<br>- €65<br>- €300 | - €50<br>- €200<br>- €350 | - €50<br>- €175<br>- €300 | - €50<br>- €225<br>- €400 | - €80<br>- €115<br>- €150 | - €100<br>- €150<br>- €200 |
| **Comfort level** | 0.5 (0.25) | - 1st class<br>- 2nd class | | - Business<br>- Economy | | / | |
| **Perceived risk of infection with covid-19** | -0.05 (0.025) | - 1 (very low)<br>- 3 (medium)<br>- 5 (very high) | | - 1 (very low)<br>- 3 (medium)<br>- 5 (very high) | | *1 (very low)* | |

*Additional questions*

In addition to the rating and bridging experiments, respondents were presented with travel-related and socio-demographic questions. To get a better idea of respondents' long-distance travel characteristics, we asked them (1) how many times they had travelled to European destinations in 2021, (2) what was the most frequent purpose of those trips, (3) who paid for those trips and (4) who they travelled with. We also asked them to state their preferred travel mode for both the shorter (~500km) and longer (~1000km) context trips. As the Omicron variant had just become the dominant strain of covid-19 a month before the survey took place, we included a question on the perception thereof; whether they are more, equally or less worried about it, as compared with the previously dominant Delta variant. Regarding the socio-demographic information, respondents were asked to elicit their age, gender, income, completed level of education, working status, household composition and access to a car.

## 2.2 Model estimation

The separate rating and bridging experiments, forming the complete HII dataset, are modelled separately. In both Conventional HII and the HII variant, rating experiments are modelled by means of a multiple linear regression and the bridging experiment as a discrete choice model (DCM). As mentioned in the introduction, we apply an LCCM to account for respondent heterogeneity in the bridging experiment. By doing so, we obtain information on the probability of each individual to belong to a certain latent group in the population, which is subsequently used to estimate a weighted least squares (WLS) regression for the rating experiment. As the class allocation probabilities are a prerequisite for WLS regression, we firstly explain the LCCM estimation for the bridging experiment, before proceeding with the regression analysis for the rating experiment. The choice model bridging experiment is modelled with the help of the PandasBiogeme python package (Bierlaire, 2020) and the rating experiment with IBM SPSS Statistics (Version 26).

*Bridging experiment: Latent class choice model*

The bridging experiment is a discrete choice experiment and is therefore modelled using a DCM. The decision rule that respondents used to make their decisions is assumed to be utility maximisation (McFadden, 1974). As a point of departure, different MNL models are estimated, testing for different parameter specifications, capturing potential interaction and non-linear effects. The model is then extended to also capture respondent heterogeneity, which gives more detailed insights into individuals' choice behaviour.

As outlined in the Introduction section, several different DCM specifications exist that are able to capture respondent heterogeneity. Two of the most prominent are the ML and LCCM approaches (Greene & Hensher, 2003), with so far only the former being applied in HII experiments. The benefits of ML models is that they are able to capture heterogeneity with a fairly small number of parameters, making them very parsimonious in the estimation. For attributes deemed to vary in the population, a second parameter (standard error) is estimated, giving information on the width of the (normal) distribution of the attribute's perception in the population. In HII experiments, the perception of the rating attribute is then linked with a regression analysis, meaning that a single (normally distributed) parameter is a good way of achieving this.

On the other hand, LCCMs capture heterogeneity by estimating several distinct MNL models, to which individuals are allocated to in a probabilistic fashion, based on the likelihood of their observed choices (Equation 1). In this manner, each MNL model represents a distinct class or segment of the population, making their interpretation very straightforward. Another benefit of LCCMs is that the class allocation function, used to classify individuals, can be extended with socio-demographic information, providing valuable insight into the composition of each class. Specifically, this information can be included in the class allocation utility $C_{n,s}$ (Equation 2). The class allocation probability $\pi_{n,s}$ is then carried on into the WLS regression, to model the rating experiment, which is further elaborated on in the following section.

*Equation 1. Formulation of the LC model*

$$P_n(i|\beta) = \sum_{s=1}^{S} \pi_{n,s} \cdot P_n(i|\beta_s)$$

where:

$P_n(i|\beta_s)$  Choice probability of respondent *n* selecting alternative *i*, given a set of parameters $\beta_s$

$\pi_{n,s}$  Probability of respondent *n* belonging to class *s*

*Equation 2. Formulation of the class allocation probability*

$$\pi_{n,s} = \frac{e^{C_{n,s}}}{\sum_{l=1}^{S} e^{C_{n,l}}}$$

where:

$C_{n,s}$  Utility of respondent *n* belonging to class *s*

*Rating experiment: Regression*

The results of the rating experiments in HII are analysed using a multiple linear regression approach, to capture the impact of each individual aspect onto the scoring of the attribute. As we aim to capture the different underlying perceptions of risk by different population segments, we propose to estimate *S* number of regression models, one for each segment obtained from the LCCM. To differentiate between the models, we apply a WLS regression, as opposed to the ordinary least squares (OLS) regression. In OLS, each data point contributes equally to the estimation of the regression model. WLS is a generalisation of OLS, wherein a weight is added for each individual data point, indicating the trust or accuracy of the researcher into that specific data point. The weight is an additional parameter in the regression function that takes a values between 0 and 1, indicating how much it should contribute to the model estimation. When all data points have a weight of 1, the WLS reduces to an OLS regression.

This approach provides a great opportunity to estimate separate regression models for each individual population segment. LCCMs allow for the calculation of a class allocation probability for each individual, which can be used as the weight in WLS. Based on this, the WLS formulation is adapted as shown in Equation 3.

*Equation 3. Formulation of the class allocation probability*

$$WSS_s = \sum_{n=1}^{N} \left( \pi_{n,s} \cdot \left( y_n - \sum_{k=1}^{K} x_{n,k} \cdot \beta_{s,k} \right)^2 \right)$$

where:

| | |
|---|---|
| $WSS_s$ | Weighted sum of squares for segment *s* |
| $y_n$ | Observed value of perceived risk for respondent *n* |
| $x_{n,k}$ | Attribute level of attribute *k*, evaluated by respondent *n* |
| $\beta_{s,k}$ | Parameter capturing the sensitivity of parameter *k* in segment *s* |

## 2.3   Data collection

The survey was administered to the respondents of the panel managed by the Dutch Railways (NS) (NS, 2020). In total, 938 responses are obtained between 8th of February and 8th of March 2022. This data is filtered based on a minimal response time of five minutes and maximum of 30 minutes, resulting in 705 fully valid responses. A lower boundary is set to remove speed runners from the data. As the connection between the rating and bridging experiments is crucial, an upper boundary is also set for the response time, to guarantee that respondents are still conscious of this connection.

The sample's socio-demographics characteristics are compared to those of the Dutch population and presented in Table 3. The sample is skewed with regard to the overall Dutch population, consisting of an above average share of older individuals and therefore a larger share pensioners. Additionally, the sample has a higher than average level of education, with 63% having bachelor's degree or higher. While it can be seen that the sample is not representative of the Dutch population, we cannot be certain how representative it is of the Dutch long-distance-travelling public, our target population. The respondents are members of the Dutch Railways panel, indicating a potential preference towards train travel. However, as regular train travel is not a prerequisite for joining the panel, this may not necessarily be the case. The impact of the sample characteristics on the model results are examined in the Conclusion & Discussion section.

*Table 3. Sample and population socio-demographic characteristics*

| | | Population | Sample |
|---|---|---|---|
| Gender | Female | 50 % | 49 % |

|  | Male | 50 % | 50 % |
|  | other |  | 2 % |
| Age | 18-34 | 27 % | 12 % |
|  | 35-49 | 22 % | 13 % |
|  | 50-64 | 26 % | 27 % |
|  | 65+ | 25 % | 48 % |
| Education | Low | 29 % | 14 % |
|  | Middle | 36 % | 23 % |
|  | High | 35 % | 63 % |
| Income | Below average | 40 % | 21 % |
|  | Average | 52 % | 50 % |
|  | Above average | 8 % | 8 % |
|  | Did not say |  | 21 % |
| Working status | Working | 66 % | 45 % |
|  | Retired | 23 % | 42 % |
|  | other | 11 % | 13 % |

## 2.4 Covid-19 situation and survey context

With the survey being undertaken during the covid-19 epidemic and various government measures in place, it is important to understand the context under which the respondents have been answering the survey. An overview of the situation and implemented measures can be seen in Figure 1. At the start of the year in 2022, around 70% of the Dutch population was fully vaccinated and half have also received their booster shot (Rijksoverheid, 2022b). The first cases of the Omicron variant of covid-19 had been diagnosed in the Netherlands in mid-November 2021 (Seveno, 2021), with the government announcing new lockdown measures not long after, on December 18th. Through the course of January 2022, the Omicron variant of covid-19 became dominant in the Netherlands, representing 47% on the 3rd and reaching 98% by the 31st of January. At the same time, with hospitalisations and ICU admissions declining, the most restrictive measures had been lifted, namely the reopening of schools and (non-essential) shops on January 15th. This was followed by the reopening of bars and restaurants on January 26th, although a proof of vaccination, recovery or testing was still required and special occupancy limits were still in place, to comply with social distancing norms. Halfway through the survey collection stage, on February 25th, a mask mandate in public indoor areas and social distancing norms (1.5m distance) had been lifted, with the exception of masks on PT (Rijksoverheid, 2022a). Mask were no longer required on PT as of March 23rd, with widespread testing of the population ending on April 11th (NOS Nieuws, 2022).

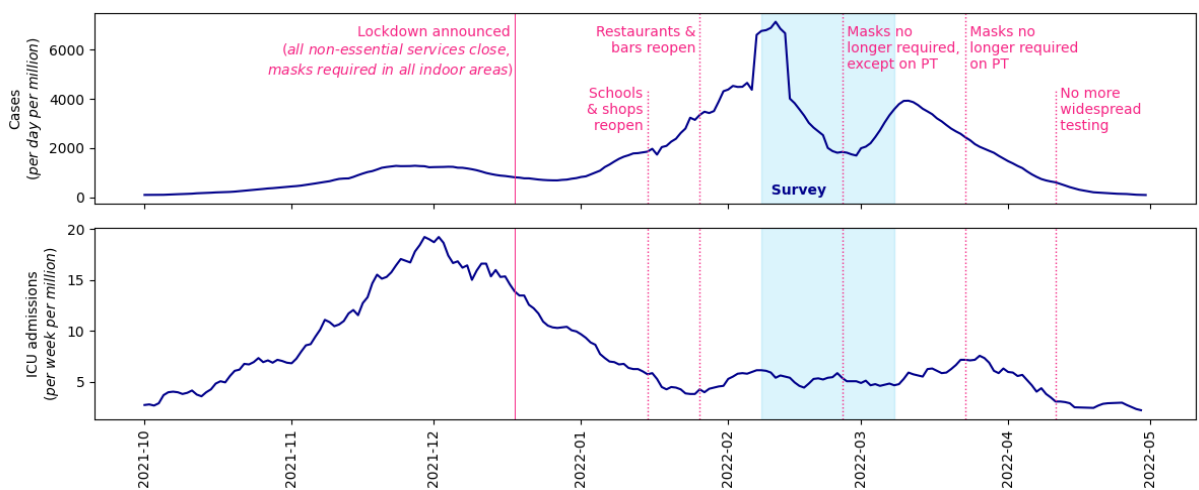

*Figure 1. Covid-19 situation (cases and ICU admittance) in the Netherlands, from October 2021 until May 2022. The data collection period is indicated by the light blue shade.*

# 3 Results

From the bridging experiment, we obtain a 4-class latent class model, with a loglikelihood of -4,030, yielding an adjusted rho-squared of 0.34. Using the sample segmentation, we estimate four weighted least squares regression models – one per market segment – which achieve $R^2$ values ranging from 11.3% to 15.9%. The full model outcomes, with parameter estimates and t-values are shown in Table 5 and Table 6 for the bridging experiment and rating experiments, respectively.

For the bridging experiment, multiple model specifications are tested to obtain more detailed information, particularly for the infection risk perception attribute. The final model is specified with mode-specific risk perception parameters (for air and train, the risk for car is fixed to 1 – low). Each mode is also associated with two separate risk parameters: (1) a fixed penalty that is associated with its respective mode and (2) and a time-based penalty. This allows us to test if individuals' (and different segments) perceived risk is time-dependent or not. In other words, do people feel less safe if they spend a longer time in a higher-risk situation or not. Our modelling approach also allows us to capture situations where travellers assign both a fixed and a time-based penalty to a travel mode. To retain consistency between the rating and bridging experiments, we treat perceived risk as an interval variable. We refer the interested reader to Heufke Kantelaar et al. (2022) for a detailed deliberation on why this is preferable. Considering the comfort attribute, we apply the same "fixed and time-based penalty" approach, but keep it generic across the rail and air modes. Travel time and cost are also assumed to be generic and thus perceived equally for all three modes.

Estimating the LCCM models, we start by determining the optimal number of segments by means of a static class membership function (constant only) (Hess et al., 2008). In order to determine the most suitable number of classes for the latent class choice model, we consider several criteria (Table 4). Weighing the model fit against the number of estimated parameters, we calculate the adjusted rho-squared and BIC, where the former should be as high as possible, whereas the latter as low as possible. For both, we see that a model with seven segments is the best performing one. The size of the segments should also be sufficiently large, to guarantee a meaningful representation of the segment. For this, we apply a rule-of-thumb size of at least 10% of the sample for the smallest segment. Applying this rule, results in excluding models of six segments or more. Finally, we consider interpretability and the number of significant parameters. With the latter, we observe hardly any change between the four and five segment models. Additionally, in the five segment model, two segments are nearly identical in terms of their parameter estimates. We therefore choose to continue our analysis with an LCCM with four specified segments.

All socio-demographic information is incorporated into the class membership function, to improve the predictive capabilities of the model. If the parameters for all three segments (one segment serves as the baseline) are insignificant ($p > 0.1$), the socio-demographic is removed. This is done iteratively, until only significant socio-demographics remain in the class membership function.

*Table 4. Number of segments and their respective model fits and characteristics*

| Number of segments | Adjusted Rho-squared | BIC | Number of significant parameters ($p < 0.05$) | Size of the smallest segment |
|---|---|---|---|---|
| 1 | 0.1798 | 10,230 | 7 (70%) | 100% |
| 2 | 0.2612 | 9,252 | 10 (48%) | 49% |
| 3 | 0.2975 | 8,851 | 15 (47%) | 16% |
| 4 | 0.3131 | 8,709 | 19 (44%) | 16% |
| 5 | 0.3217 | 8,652 | 20 (37%) | 12% |
| 6 | 0.3286 | 8,617 | 21 (32%) | 8% |

| 7 | 0.3346 | 8,592 | 26 (34%) | 6% |

In the regression models, capturing the infection risk perception, all attributes are dummy coded. For nominal and ordinal attributes, this (or effects coding) is the only option. For the three ratio attributes (share of occupied seats, number of daily infections and vaccination rate), we also apply dummy coding to capture any potential non-linear effects that may be present in the perception of risk.

In the following sections, each of the four segments is described, based on their travel preferences, infection risk perception and socio-demographic characteristics. The implications of their preferences on modal split are then discussed in the following section, by means of a sensitivity analysis. The four segments are:

- Segment 1: Time-sensitive travellers
- Segment 2: Prudent travellers
- Segment 3: Frequent train-loving travellers
- Segment 4: Cautious car travellers

### Segment 1: Time-sensitive travellers

Time-sensitive travellers are, as the name implies, the most sensitive of the four segments to travel time, with a Willingness-to-Pay (WtP) of 72€/h compared to a sample average of 40€/h. Figure 2 shows the impact of perceived risk on modal preferences (fixed penalty) in the top row, whereas the bottom row displays the impact on WtP (time-based penalty). We can see that Time-sensitive travellers perceive risk as time-dependent only (sensitivity to travel time increases with risk, resulting in a higher WtP for higher risk) and do not associate any fixed penalty for risk (modal preferences are constant). Further, we observe that time sensitivity doubles for Time-sensitive travellers if the level of perceived risk jumps from low (level 1) to high (level 5). Mode-specific constants indicate a strong preference for train and the largest aversion towards the private car (all else being equal) amongst all segments. Comfort (both risk and time-dependent) is insignificant for this segment.

Turning to the perception of infection risk, members of this segment seem to be strongly influenced by the infection and vaccination rates, as well as the mask requirements. Figure 3 shows the impact of the different attributes on perceived risk. Official advice, like government travel advise and entry requirements do not have a strong influence on the perceived risk of Time-sensitive travellers. This seems to indicate that they prefer making their own informed decision rather than rely on the government travel advice. Curiously, all segments associate yellow advice with lower risk and red with higher risk, compared to the green. A possible explanation for this could be that people believe yellow (and orange) advise will deter enough people from travelling to that destination to make it somewhat safer, offsetting the possible higher risk which substantiated the advice. A red advice on the other hand seems to indicate to members of all segments that the risks are simply too great.

Members of this segment are quite representative of the sample as a whole in terms of income, education and gender. They have a fairly low car ownership (only 50% have their own) and are the least frequent long-distance travellers of the four, with 68% not making a single long-distance trip in 2021, compared to the 61% sample average. Those who did travel, travel above average for work, indicating that they primarily travel only when they have to (limited leisure trips). As for the stated modal preferences, they prefer train or air for shorter and only air for longer trips. This last can also be observed in the ternary chart in Figure 4. The ternary chart indicates the market share for each of the three available modes. The altitude of the triangle indicates the market share; the closer to the corner, the higher the market share for each of the three modes and their corresponding corners. The preference based on the ASCs is determined assuming all other attributes are equal / zero (i.e. ceteris paribus), and only ASCs enter the utility function.

*Table 5. Model fit, parameter estimates and class allocation parameters of the mode choice model*

| | Model fit | | | | | | | |
|---|---|---|---|---|---|---|---|---|
| Null LL | -6,196 | | | | | | | |
| Final LL | -4,030 | | | | | | | |
| Adj. Rho-square | 0.34 | | | | | | | |
| BIC | 8499 | | | | | | | |

| | Taste parameters | | | | | | | |
|---|---|---|---|---|---|---|---|---|
| | Segment 1 | | Segment 2 | | Segment 3 | | Segment 4 | |
| | Est | t-val | Est | t-val | Est | t-val | Est | t-val |
| *Constants* | | | | | | | | |
| Air | 1.3500 | 3.59 *** | -1.1500 | -2.74 *** | -2.2500 | -2.97 *** | -3.3500 | -4.24 *** |
| Train | 2.7300 | 3.89 *** | 0.0592 | 0.21 | 2.5600 | 6.02 *** | -0.0568 | -0.13 |
| *Common parameters* | | | | | | | | |
| Cost | -0.4500 | -6.89 *** | -0.6120 | -8.36 *** | -0.5330 | -5.22 *** | -0.9220 | -7.43 *** |
| Travel time | -0.3250 | -3.92 *** | -0.3100 | -7.03 *** | -0.2000 | -4.18 *** | -0.1400 | -3.16 *** |
| Comfort [baseline] | | | | | | | | |
| Comfort [time-based] | 0.0411 | 0.72 | 0.0867 | 2.64 *** | 0.1040 | 1.43 * | | |
| *Risk parameters* | | | | | | | | |
| Train   Baseline | | | -0.6990 | -2.86 *** | -0.5240 | -3.15 *** | -0.2700 | -1.34 * |
| Train   Time-based | -0.0767 | -4.50 *** | -0.0015 | -0.07 | | | -0.0620 | -3.04 *** |
| Air     Baseline | -0.0113 | -0.06 | -0.3480 | -1.57 * | -0.1070 | -0.14 | | |
| Air     Time-based | | | | | | | -0.0344 | -0.47 |

| | Class allocation parameters | | | | | | | |
|---|---|---|---|---|---|---|---|---|
| Constant | -1.3600 | -1.79 ** | | | -1.4500 | -1.56 * | -0.4090 | -0.43 |
| Age | 0.0249 | 2.66 *** | | | 0.0283 | 2.37 ** | 0.0140 | 1.29 * |
| Car ownership | -0.7260 | -2.03 ** | | | -1.3600 | -3.61 *** | -0.1550 | -0.45 |
| Travel frequency | -0.1800 | -0.93 | *Baseline* | | 0.1850 | 1.00 | -0.3700 | -1.84 * |
| Female | 0.5350 | 2.22 ** | | | 0.2490 | 0.76 | 0.2570 | 0.92 |
| Air for short trips | 0.2750 | 0.61 | | | -7.5300 | -7.55 *** | 0.6630 | 0.65 |
| Car for short trips | -1.4600 | -3.65 *** | | | -1.9400 | -2.79 *** | 0.4860 | 1.06 |
| Air for long trips | 0.2320 | 0.51 | | | -4.4700 | -1.09 | -2.6300 | -5.53 *** |
| Train for long trips | 0.1930 | 0.40 | | | 1.1400 | 2.41 ** | -0.4030 | -0.90 |

*\*\*\* p ≤ 0.01, \*\* p ≤ 0.05, \* p ≤ 0.2*

Segment 2: Prudent travellers

Similar to Time-sensitive travellers, the Prudent travellers are also quite sensitive to time, but slightly less so, with a WtP of 50€/h. Contrary to the previous segment, Prudent travellers perceive risk primarily as both mode-dependent and time-independent, with train and car being equally preferred at low risk, with train decreasing rapidly in preference to the point where even an aircraft is preferred over a train when the perceived level of risk is very high. They are also willing to pay for a class upgrade: based on the duration of their travel, they are willing to pay an additional 14€/h to travel in first/business class.

Similar to the Time-sensitive travellers, Prudent travellers are also strongly influenced by the infection rate, vaccination rate and masking requirements. While government travel advise is also less relevant, they do consider the entry requirements more than other classes do, namely more stringent regulations make them feel safer.

Describing the demographics of this segment, Prudent travellers tend to be the youngest and most male dominant (59%) of the segments. Members of this class are also most likely to be employed and least likely to be retired of any segment. In 2021, they travelled slightly more than the average survey respondent, with 10% making four or more long-distance trips. In terms of their modal preference (Figure 4), they prefer using their car for shorter trips and flying for longer trips.

*Table 6. Model fit and parameter estimates of the risk perception regression models*

| | Model fit | | | | | | | |
|---|---|---|---|---|---|---|---|---|
| | Segment1 | | Segment2 | | Segment3 | | Segment4 | |
| $R^2$ | 14.02% | | 15.85% | | 12.04% | | 11.30% | |
| BIC | 10,751 | | 10,635 | | 19,300 | | 11,686 | |

| | Parameter estimates | | | | | | | |
|---|---|---|---|---|---|---|---|---|
| | Est | t-val | Est | t-val | Est | t-val | Est | t-val |
| Constant | 3.2769 | 54.68 *** | 3.2481 | 48.75 *** | 3.2085 | 49.78 *** | 3.2048 | 56.97 *** |
| ***No mask required*** | | | | | | | | |
| Any mask | | | 0.1752 | 2.20 ** | | | | |
| Surgical mask | -0.3019 | -2.44 ** | | | | | | |
| FFP2 mask | -0.3439 | -4.44 *** | -0.3618 | -5.24 *** | -0.2499 | -4.91 *** | -0.2424 | -5.00 *** |
| ***Status quo*** | | | | | | | | |
| Increased | 0.3252 | 3.40 *** | 0.4331 | 4.24 *** | 0.2436 | 3.34 *** | 0.2387 | 4.81 *** |
| Weekly disinfection | 0.2315 | 2.61 *** | 0.4534 | 4.82 *** | 0.3296 | 4.67 *** | 0.2444 | 4.95 *** |
| Daily disinfection | | | | | | | | |
| ***Nothing*** | | | | | | | | |
| Ventilation only | | | | | | | | |
| AC w/o HEPA filters | -0.2530 | -5.19 *** | -0.3550 | -5.52 *** | -0.3236 | -5.15 *** | -0.3773 | -7.28 *** |
| AC w/ HEPA filters | -0.2997 | -6.47 *** | -0.2974 | -6.21 *** | -0.1343 | -2.60 *** | -0.2519 | -5.79 *** |
| ***None*** | | | | | | | | |
| 3G | | | | | 0.1953 | 2.38 ** | | |
| 2G | | | -0.2586 | -3.56 *** | -0.1903 | -3.42 *** | -0.1265 | -2.50 ** |
| 2G+ | -0.2073 | -4.60 *** | -0.2923 | -5.60 *** | | | -0.1113 | -2.49 ** |
| ***Green advice*** | | | | | | | | |
| Yellow advice | -0.2240 | -2.37 ** | -0.1639 | -2.03 ** | -0.3295 | -3.54 *** | | |
| Orange advice | | | | | -0.4432 | -4.64 *** | | |
| Red advice | 0.3667 | 3.79 *** | 0.4587 | 5.04 *** | 0.3791 | 6.60 *** | 0.5234 | 10.65 *** |
| ***25% seats full*** | | | | | | | | |
| 50% seats full | | | 0.1951 | 2.25 ** | | | | |
| 75% seats full | | | | | -0.2507 | -4.31 *** | | |
| 100% seats full | 0.2707 | 5.31 *** | 0.3594 | 7.38 *** | 0.6007 | 9.28 *** | 0.3139 | 6.27 *** |
| ***100 cases*** | | | | | | | | |
| 10.000 cases | 0.3878 | 3.17 *** | | | | | | |
| 25.000 cases | 0.5764 | 7.42 *** | 0.4363 | 7.53 *** | 0.1511 | 2.43 ** | 0.3105 | 5.85 *** |
| 100.000 cases | 0.3078 | 4.68 *** | 0.2623 | 3.06 *** | 0.2538 | 4.18 *** | 0.2723 | 5.16 *** |
| ***15% vaccinated*** | | | | | | | | |
| 30% vaccinated | -0.2331 | -3.12 *** | -0.3830 | -5.15 *** | | | -0.2234 | -3.03 *** |
| 70% vaccinated | -0.1939 | -2.06 ** | -0.3598 | -4.69 *** | | | -0.2369 | -3.08 *** |
| 90% vaccinated | -0.5056 | -6.58 *** | -0.5679 | -6.76 *** | -0.1854 | -3.09 *** | -0.2763 | -4.78 *** |

*\*\*\* p ≤ 0.01, \*\* p ≤ 0.05, \* p ≤ 0.2*
***Baseline attribute levels shown in bold italic***

## Segment 3: Frequent train-loving travellers

Contrary to the previous two segments, who exercise a considerable trade-off behaviour amongst different travel attributes, the second two segments have stronger mode-related preferences, and are thus not easily swayed to try a different travel mode. As their name implies, Train-loving travellers have a strong preference for train travel, although this diminishes in case of higher perceived risk. They are also strongly averse to flying: all else being equal, they are willing to travel 24 hours longer by train than air. According to our findings, they perceive risk primarily as time-independent and associate a fixed penalty with each mode. The parameter for comfort is borderline significant (p = 0.15), but indicates quite a high WtP for an upgrade of almost 20€/h or over 50% more compared to economy class (with a WtP for second class of 38€/h). This is the biggest relative WtP for an upgrade (the previous two segments have a relative WtP of 13% and 28% respectively).

Frequent train-loving travellers also perceive risk differently, compared to the two previously described segments. The key differentiating factor is that they seem to base their risk primarily on the government travel advice, showing the strongest sensitivity to both the yellow and orange warnings (Figure 3). Along the same lines, they have the lowest sensitivity to vaccination and infection rates. This seems to indicate that they trust the official travel advice takes this into account and they do not have to concern themselves with any additional information. They exhibit mixed behaviour when it comes to entry requirements and crowding with no clear linear trend. They are nevertheless the most sensitive in refraining from traveling in full vehicles (100% occupied).

Frequent train-loving travellers are, as one may expect, the most frequent long-distance travellers in 2021: 15% travelled four or more times. Their train-loving nature also corresponds to the lowest car ownership of any class, with only 35% having their own car and 37% having no car access at all (compared to the sample averages of 57% and 23% respectively). In terms of modal preferences, they exhibit a strong anti-flying and anti-car attitude, with most preferring to take train for short as well as long trips (Figure 4). They are the oldest, most female dominated segment (57%), having on average the highest level of education.

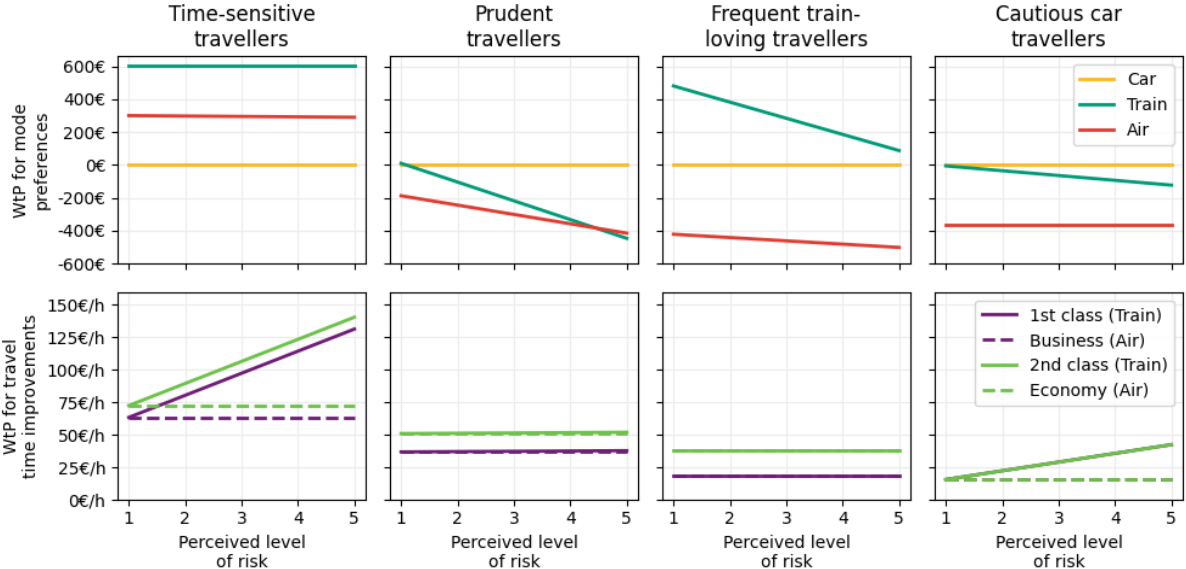

*Figure 2. Mode preferences and WtP for travel time improvements, given different levels of perceived infection risk*

## Segment 4: Cautious car travellers

The last segment we uncover has the lowest WtP towards travel time improvements: 15€/h. Interestingly, cautious car travellers seem to experience risk both as a fixed penalty and time-dependent. Although, judging based on the parameter significance, time-dependence is more important to decision-makers. In relative terms, they are the most sensitive to increasing risk on a train, as the sensitivity to travel time in high risk situations is almost three times (2.77) as high as in low risk situations. This perhaps also explains their strong preference for car, as it is perceived as safer (and was also presented as such in the survey).

Moving to what influences their perceived level of risk, they seem to exhibit the most average perception of the different attributes. However, they are the only class to not see a yellow travel warning as lowering risk. They also perceive a red warning as the most risky out of any class.

Conversely to the previous segment and analogous to their name, this segment has the highest car ownership, with 68% owning their own car and only 15% not having it at all. Their high car ownership likely influences their modal preferences, as they strongly prefer to use their car for making any kind of long-distance trips. When they travel, it is usually not alone, but with their partner or family.

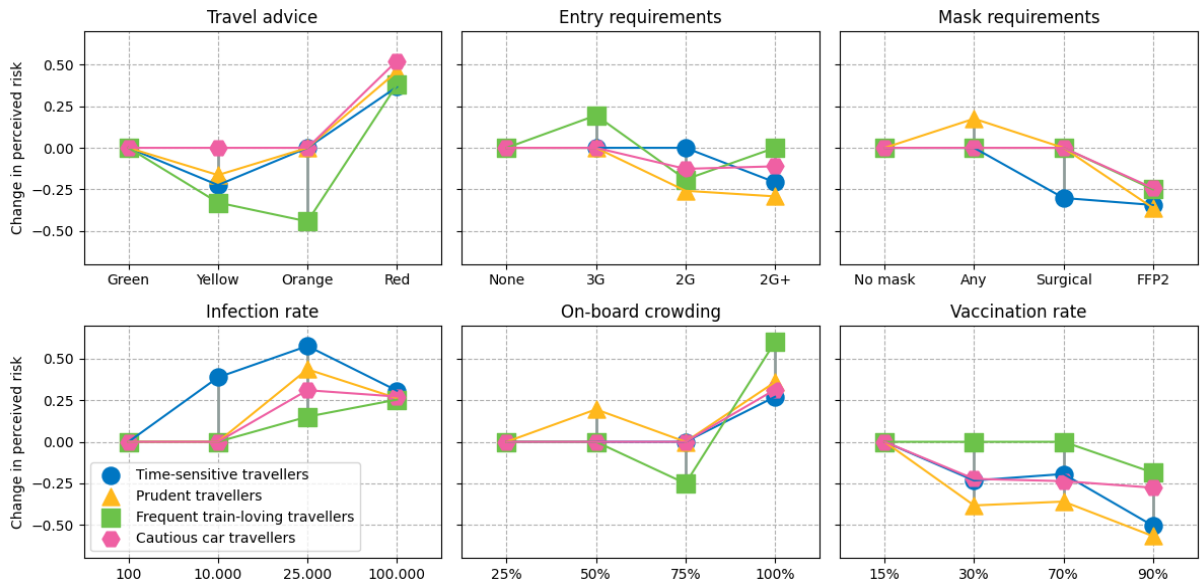

*Figure 3. Sensitivity to different risk factors for the four market segments*

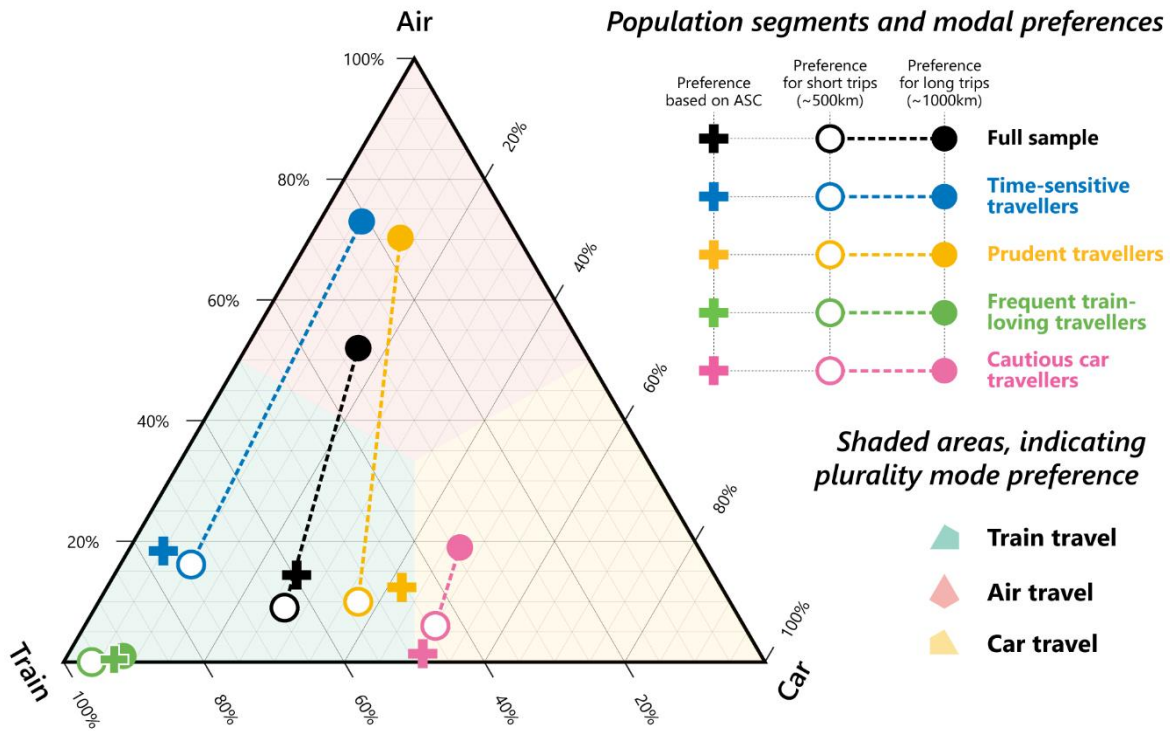

*Figure 4. Market share of different modes, based on segments' modal preferences*

## 4 Implications

To better understand how the obtained parameter estimates impact mode choice, we perform a sensitivity analysis, wherein we explore how the modal split is affected for different trip lengths and different levels of perceived risk. As the sample is not representative of the Dutch (travelling) population, the sizes of the individual segments are thus also not assumed to be representative. Hence, we do not present a joint "population" modal split in the analysis, but rather analyse changes at the level of the individual segments. The results are presented in Figure 5.

In the analysis, the level of perceived risk is only varied for the train, Levels 1 (low risk) and 3 (medium risk) and 5 (high risk). As can be seen from Figure 2, the impact of perceived risk on flying is minimal, both with respect to time and overall. We therefore decide to fix the perceived risk for flying to Level 3. Remaining consistent with our survey, the perceived risk of car is fixed to Level 1.

To obtain a value for travel time based on travel distances, we apply a linear function (Figure 6 in Appendix B) where each mode is associated with an average travel speed and a constant (intercept) that corresponds to the time spent waiting at the airport/train station. For the train, Figure 6 shows multiple different speeds; for this analysis, we assume an average speed of 160km/h, which corresponds to a partial use of high-speed rail infrastructure. For a more detailed analysis, where the variation of average train speed on market share is investigated, we refer the reader to Appendix C.

Determining price is done in a similar manner to the travel time. Based on pricing data of long-distance trips in Europe (Tanner & Provoost, 2023), we obtain a linear function for price based on distance for all three investigated modes. The functions can be seen in Appendix B in Figure 6.

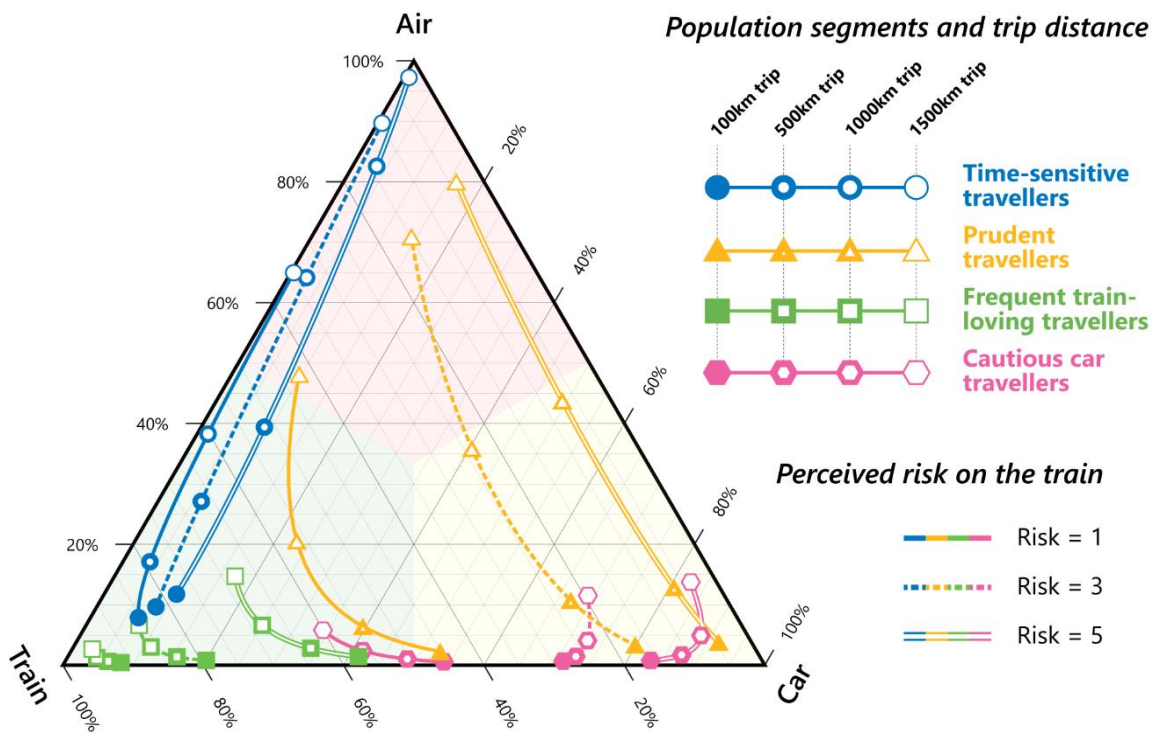

*Figure 5. Market share of different modes and the sensitivity of market segments to the variation of the level of perceived risk when travelling by train*

From the results, the first thing we notice - as is already alluded to the previous section, through the names of the segments – is that the first two segments (Time-sensitive and Prudent travellers) tend to display more trade-off behaviour, with the shifts between modes being much more pronounced. Interestingly, they each seem to have their preferred land-based mode for short-/medium-distance trips, i.e. train and car, respectively - and flying for longer trips. With respect to Time-sensitive travellers, we further observe their strong aversion to using the car, by the fact that with increasing risk on a train, they almost exclusively shift to air. For example, for a trip of ~1000km, with the risk increasing *low-medium-high*, the share of trips by train decrease *60%-30%-15%*, whereas the share of people choosing to fly increases *40%-65%-80%* and while the car sees a small increase *0%-5%-5%*.

Prudent travellers are the most mode-agnostic, exhibiting the strongest shifts between modes given the varying distance and risk. When risk is perceived low for the train (lower than for air travel), the train actually becomes the dominant mode for distances between 200km and 1,500km, with the peak observed between 500km and 1,000km. This train ridership peak also sees the biggest drop if perceived risk increases. For 500km-trips, it drops from 55% to 25% to 5% for low, medium and high risk respectively.

Moving to the second two segments, the more mode-fixed ones, they predominantly shift between land-based modes, with both largely refraining from flying: less than 15% choose it even for trips of 1,500km. For the competition between the train and car, a lower perceived risk on the train naturally leads to a higher share for the train. Interestingly, if the risk of train and car is equal, Cautious car travellers would actually use train more than car, if the trips are over 500km. However, this drops to below 30% for a medium level of risk and below 15% for high risk.

Frequent train-loving travellers seems to be least affected by this, highlighting their dedication to the train and that even very risky situations will not persuade them to shift modes, with the lowest train share of 60%, observed for short high risk trips (100km).

These results clearly indicate that making passengers feel safer on the train will result in them using it more often. And while this is an important benefit, lowering travellers' perceived risk is beneficial in its own right, as it makes the travel less stressful and more enjoyable. It may also convince more travellers to try the train. If train is perceived as less safe however, air and car travel both benefit from it, offering a modal substitution. As the sizes of the segments within the population are not known, it is difficult to determine which sees more shifting travellers. Time-sensitive travellers shift almost exclusively to air, Frequent train-loving and Cautious car travellers shift predominantly to car, whereas Prudent travellers seem to shift to both.

# 5 Conclusion & Discussion

This research investigates the subjective perception of risk, within the context of long-distance (international) travel. A Hierarchical Information Integration (HII) approach is applied to measure the perceived risk and the mode choice behaviour for trips of approximately 500km and 1,000km. The data obtained through the Dutch Railways' panel is analysed by means of a Latent Class Choice Model (LCCM). To analyse the subjective perception obtained through the rating experiment, we present a novel approach to analyse this for different population segments. We apply a Weighted Least Squares (WLS) regression model (as opposed to the commonly applied Ordinary Least Squares), and integrate therein the individual class membership probabilities obtained in the LCCM.

Based on the estimated LCCM, we uncover four distinct segments with respect to their varying travel behaviour preferences. Two segments, dubbed the *Time-sensitive travellers* and *Prudent travellers*, have an above average value of time: 72€/h and 50€/h, respectively. They also each have their preferred land-based mode for shorter trips of (train and car, respectively) and flying for longer distances. The two segments with lower values of time, the *Frequent train-loving travellers* and *Cautious car travellers*, are more tied to their respective mode of choice (as their names imply), switching to other modes only in extreme circumstances (very low/high risk and or a very long trip).

With respect to risk, some seem to perceive it as based on time (a longer exposure time results in a higher disutility), whereas others see it as a fixed penalty, dependant on the level of risk and the travel mode, but not on travel time. In cases when the risk is time-based, that is the case only for the train, but not for the flight. *Time-sensitive travellers* tend to perceive risk as time-dependent, with the difference in travel time penalty almost doubling (a 95% increase) when increasing the level of perceived risk from a low to high. *Prudent travellers* and *Frequent train-loving travellers* on the other hand tend to perceive risk as time-independent. In both segments, travellers associated a bigger penalty per train travel-minute than per air travel-minute. Finally, *Cautious car travellers* tend to see it as a mix of both, a fixed and a time-based penalty.

Surprisingly, only a single segments showed some degree of sensitivity, yet low, to risk while flying (*Prudent travellers*), that being the time-independent perception. And even then, the significance level of the parameter is quite low. One possible reasoning could be that flying is perceived equally risky, no matter the level of risk, a preference which is captured not in the risk-perception parameters, but rather by the mode-specific constant. This could also explain why air (including the time spent at airports which is an integral part of the air travel experience) had the lowest overall preference amongst all four segments.

What is considered risky tends to differ among segments, however they do generally tend to impact it in the same direction, with the main difference being in magnitude. *Time-sensitive travellers* and *Prudent travellers*, seem to prefer analysing the data themselves, being the most sensitive to infection and vaccination rates, while simultaneously being less affected by government travel advice. *Frequent train-loving travellers* and *Cautious car travellers* on the other hand, exhibit the opposite behaviour, with the official advice being an important deciding factor.

Comparing the impacts of different risk factors, it can be seen that certain policy measures (mask wearing, entry requirements etc.) can help in reducing the level of perceived risk, but their contribution to reducing perceived risk tends to be lower than the increase caused by external factors. Curiously, the impacts of cleaning and air circulation are very inconclusive. With respect to air circulation, the model shows that three of the four segments see the addition of HEPA filters as more negative. This likely stems from the fact that the majority do not know what HEPA filters are or how they work, as this was not explained during the survey. Regarding the cleaning policy, travellers' expectations of the current norm seems to be a daily disinfection, as all four segments saw no difference between the two. Furthermore, performing only weekly disinfection or merely increased cleaning of touchpoints was perceived as more risky than the status quo by all segments.

It is also important to note that risk perception and safety measures can have a profoundly different impact on travellers on long-distance trips as opposed to commute trips. From previous work, crowding is often cited as a major concern from travellers on commute trips. In our results however, this does not seem to be the case. While we can see a penalty for 100% occupancy, this is not very high, whereas occupancy levels below 100% have a mixed and negligible impact. This could potentially stem from a higher traveller density typically observed on commute trips, with people often standing. In contrast, long-distance trips tend to have more space for passengers, with people standing being a very uncommon sight. The much longer exposure time experiences on long-distance trips (up to several hours) could also contribute to a lower reduction in risk perception of safety measures, as travellers may perceive the measures not (as) effective to mitigate exposure during such a long time.

With that in mind, it is apparent that policy measures, while not fully mitigating the increase in risk due to high infection numbers, do still help in making travellers feel more at ease. It is important to clarify however, that while some policies may reduce the perceived risk of individuals, these do not necessarily coincide with measures which tend to be perceived as actually reducing the rate of transmission. There may be measures which make people feel safer, but have no noteworthy impact on the viral transmission, whereas others may not be perceived as relevant at all, they may have a significant role in reducing transmission. And, as can also be observed from our results, the perception of measures also varies between people. Hence, a measured and balanced combination of measures is likely needed.

The Weighted Least Squares regression proved to be a good approach to distinguish the perception of risk among different travellers. Especially for factors such as infection, vaccination and travel advice, we see a clear trend for each of the segments. Nevertheless, a key outlook for future research is to develop a model, which is capable of segmenting the sample based on both the rating and bridging experiments simultaneously. In our study, this is done purely on the bridging experiment, meaning that differences in travel behaviour are the dominant driving force behind the segmentation, and the key differentiator among travellers. If segmentation would be done exclusively on the rating experiment or ideally on both, we would expect to see a more pronounced difference in risk perception as well.

Our results are based on a sample that is obtained through the Dutch railways' panel which might be representative of the travelling population, however, this cannot be empirically verified. With that in mind, it is important to not generalise the conclusions of our study onto the population. By estimating an LCCM, this issue can be somewhat mitigated, as each individual group of travellers is allocated to a different MNL model. We therefore do not make any claims in regard to the sizes of the segments in the sample. Given our fairly large sample, and the decent representation of all groups, we believe however that we do capture the key traveller segments present in the population while their shares remain unknown.

Additional studies on long-distance travel and safety and risk perception, in different contexts and trip purposes, should further expand on the findings from our work. A dataset based on a representative sample of the population could offer information on the sizes of different traveller segments as well, to benchmark against our results. Additionally, including an opt-out option would also allow concluding on what is the tolerance (for price, time and risk) that travellers are willing to accept on such trips, as they tend to be less essential than commute trips: for leisure trips for example, the decision order of choosing the destination first and then the mode/route is likely to be inverted, and thus the threshold to opt-out may be lower. Business travel, while much less flexible in the past, has also seen a fundamental shake-up with online meeting becoming the norm.

## Acknowledgement

We would like to thank the respondents of the Dutch railways panel for their time and effort and providing us with their decision-making preferences for long-distance travel in the light of COVID-19.

This research was also supported by the CriticalMaaS project (grant number 804469), which is financed by the European Research Council and Amsterdam Institute for Advanced Metropolitan Solutions.

# Appendix A

Survey designs are constructed utilising Ngene software (ChoiceMetrics, 2018). Below, the design for the rating experiment (Table 7), shorter context (~500km, shown in Table 8) and longer context (~1000km, show in Table 9) bridging experiment designs are presented.

*Table 7. Rating experiment design*

| Task | On-board crowding | Face mask policy | Air circulation | Cleaning policy | Travel advice | Entry requirements | Infection rate | Vaccination rate | Block |
|---|---|---|---|---|---|---|---|---|---|
| 1 | 25 | 0 | 0 | 0 | 0 | 0 | 0.1 | 15 | 3 |
| 2 | 75 | 0 | 3 | 2 | 3 | 3 | 25 | 30 | 4 |
| 3 | 25 | 2 | 3 | 1 | 2 | 2 | 100 | 70 | 4 |
| 4 | 75 | 1 | 0 | 3 | 0 | 3 | 100 | 70 | 4 |
| 5 | 50 | 3 | 3 | 0 | 1 | 1 | 25 | 70 | 1 |
| 6 | 50 | 0 | 1 | 3 | 3 | 2 | 0.1 | 70 | 1 |
| 7 | 25 | 1 | 2 | 1 | 3 | 0 | 100 | 30 | 3 |
| 8 | 50 | 1 | 0 | 0 | 3 | 3 | 10 | 90 | 1 |
| 9 | 100 | 0 | 1 | 0 | 2 | 1 | 100 | 15 | 2 |
| 10 | 75 | 2 | 1 | 1 | 1 | 1 | 10 | 90 | 3 |
| 11 | 25 | 3 | 1 | 3 | 2 | 1 | 0.1 | 30 | 4 |
| 12 | 100 | 1 | 3 | 3 | 1 | 0 | 10 | 15 | 2 |
| 13 | 100 | 3 | 2 | 1 | 2 | 3 | 0.1 | 15 | 2 |
| 14 | 50 | 3 | 1 | 2 | 0 | 2 | 100 | 30 | 1 |
| 15 | 50 | 0 | 2 | 3 | 0 | 1 | 25 | 90 | 1 |
| 16 | 75 | 1 | 3 | 0 | 0 | 2 | 0.1 | 90 | 4 |
| 17 | 100 | 2 | 0 | 1 | 1 | 2 | 25 | 30 | 3 |
| 18 | 75 | 3 | 0 | 2 | 3 | 0 | 25 | 70 | 3 |
| 19 | 100 | 2 | 2 | 2 | 2 | 0 | 10 | 90 | 2 |
| 20 | 25 | 2 | 2 | 2 | 1 | 3 | 10 | 15 | 2 |

*Table 8. Bridging experiment – Shorter trip design (~500km)*

| | Train | | | | Aircraft | | | | Car | | |
|---|---|---|---|---|---|---|---|---|---|---|---|
| Task | Travel time | Travel cost | Comfort | Risk | Travel time | Travel cost | Comfort | Risk | Travel time | Travel cost | Block |
| 1 | 6 | 30 | 0 | 5 | 3 | 50 | 1 | 1 | 6.5 | 150 | 1 |
| 2 | 4.5 | 300 | 0 | 3 | 3 | 175 | 1 | 1 | 8.5 | 80 | 3 |
| 3 | 4.5 | 300 | 0 | 3 | 5 | 175 | 1 | 5 | 4.5 | 80 | 1 |
| 4 | 4.5 | 30 | 1 | 3 | 4 | 175 | 0 | 3 | 6.5 | 115 | 3 |
| 5 | 3 | 300 | 1 | 1 | 3 | 175 | 0 | 5 | 8.5 | 80 | 1 |
| 6 | 3 | 165 | 1 | 1 | 5 | 50 | 0 | 5 | 6.5 | 150 | 2 |
| 7 | 6 | 30 | 1 | 1 | 3 | 50 | 0 | 5 | 6.5 | 150 | 2 |
| 8 | 3 | 165 | 0 | 5 | 4 | 300 | 1 | 1 | 8.5 | 115 | 2 |
| 9 | 4.5 | 300 | 0 | 3 | 5 | 50 | 1 | 3 | 4.5 | 150 | 3 |
| 10 | 6 | 165 | 1 | 5 | 4 | 300 | 0 | 1 | 4.5 | 115 | 1 |
| 11 | 6 | 165 | 1 | 5 | 4 | 300 | 0 | 3 | 4.5 | 80 | 3 |
| 12 | 3 | 30 | 0 | 1 | 5 | 300 | 1 | 3 | 8.5 | 115 | 2 |

*Table 9. Bridging experiment – Longer trip design (~1000km)*

| | Train | | | | Aircraft | | | | Car | | |
|---|---|---|---|---|---|---|---|---|---|---|---|
| Task | Travel time | Travel cost | Comfort | Risk | Travel time | Travel cost | Comfort | Risk | Travel time | Travel cost | Block |
| 1 | 12 | 200 | 1 | 5 | 5 | 400 | 0 | 3 | 10 | 200 | 2 |
| 2 | 9 | 350 | 0 | 3 | 4 | 225 | 1 | 1 | 16 | 100 | 3 |
| 3 | 12 | 50 | 1 | 5 | 4 | 50 | 0 | 1 | 13 | 200 | 2 |
| 4 | 6 | 350 | 0 | 1 | 5 | 225 | 1 | 5 | 16 | 100 | 1 |
| 5 | 6 | 200 | 1 | 1 | 6 | 225 | 0 | 5 | 16 | 150 | 3 |
| 6 | 12 | 200 | 1 | 3 | 6 | 400 | 0 | 3 | 10 | 100 | 1 |

| 7  | 9  | 50  | 0 | 5 | 5 | 50  | 1 | 1 | 13 | 150 | 1 |
| 8  | 9  | 50  | 0 | 5 | 5 | 400 | 1 | 1 | 10 | 150 | 2 |
| 9  | 12 | 50  | 1 | 1 | 4 | 50  | 0 | 5 | 10 | 200 | 3 |
| 10 | 6  | 350 | 0 | 3 | 6 | 400 | 1 | 3 | 13 | 100 | 3 |
| 11 | 6  | 200 | 0 | 3 | 6 | 50  | 1 | 3 | 13 | 150 | 2 |
| 12 | 9  | 350 | 1 | 1 | 4 | 225 | 0 | 5 | 16 | 200 | 1 |

# Appendix B

In order to calculate the travel time and travel cost based on the distance, linear functions with an intercept are constructed. For travel time, average speeds of 100, 160 and 800km/h are chosen for the car, train and aircraft respectively. The intercepts is set to 0, 1 and 3 hours respectively. These capture the time travellers spend at the airport/train station and the slower access/egress times, whereas for car, it is assumed that no such waiting takes place. The relation between travel time, distance and speed can be seen in Figure 6. In an additional analysis (Appendix C), train speed is also varied between 70km/h and 300km/h.

With respect to travel cost, a regression analysis is performed on pricing data scrapped from the web (Tanner & Provoost, 2023). This is then used to construct a linear function with the estimated slope and intercept. Curiously, air fares for European flights are found to be fully distance independent, with an average price of €135. Train and car on the other hand seem to have an identical slope of ~€0.12 per kilometre (or 8.33km per €1), with an intercept of €8 and €5 respectively.

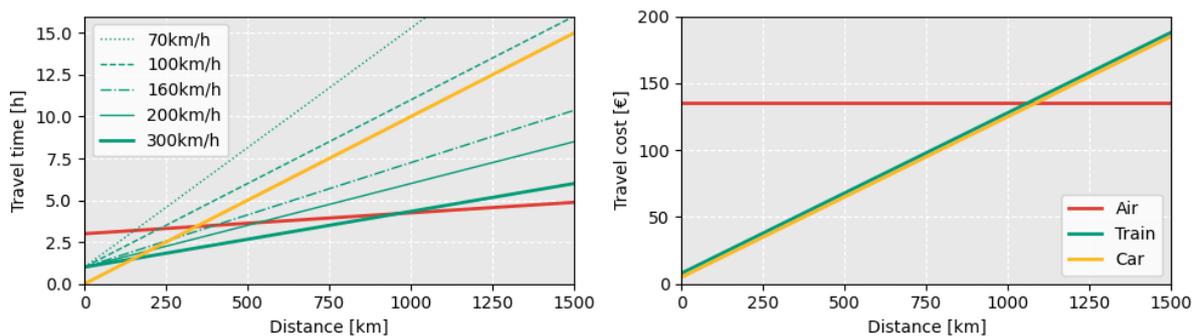

*Figure 6. Calculation of trip travel time and travel costs for different modes*

# Appendix C

Unlike car or air travel, which tend to have much more consistent average speeds, due to infrastructure between large cities being roughly at the same level, this cannot be said for trains. City pairs connected by high-speed rail (300km/h) are virtually incomparable to pairs connected with conventional rail. In the main sensitivity analysis, we assumed an average speed of 160km/h, which implies that there is some high-speed infrastructure along the way, but not the entire journey. Here, instead of varying perceived risk, we vary the average train speed, to better understand the added value of building out high-speed rail infrastructure and the potential it holds for modal shift.

What we can observe for all four segments (Figure 7) is that the difference between average speeds of 70km/h and 300km/h, over a 1,000km distance can be as much as 50 percentage points. In general, the implications of higher train speeds are largely equal to lower perceived risk (see Implications section), with Time-sensitive travellers mainly switching from air to train, Frequent train-loving and Cautious car travellers shifting from car and the Prudent travellers substituting both. We also see the same modal preference pattern, with Time-sensitive and Prudent travellers each having their own preferred land-based mode for shorter trips and flying over longer distances, whereas Frequent train-loving and Cautious car travellers predominantly stick to their preferred land-based mode (respectively to their names), even for distances of over 1,000km.

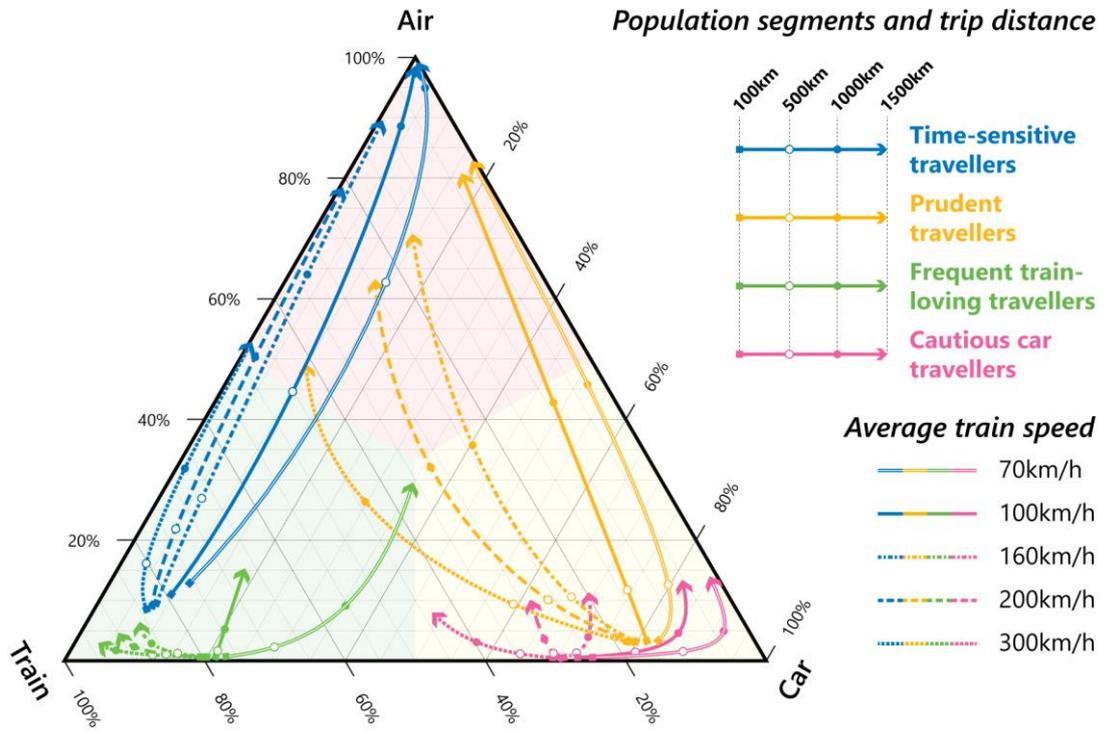

*Figure 7. Market share of different modes, for different market segments, based on average train travel speeds*